\documentclass[noshowpacs,amsmath,twocolumn,
aps,prb]
{revtex4-1}

\usepackage{setspace}
\usepackage{amsmath}
\usepackage{graphicx}
\usepackage[nearskip,margin = 0pt]{subfig}
\usepackage{subfig}

\usepackage{verbatim}
\usepackage{amsfonts}
\usepackage{amssymb}
\usepackage{epstopdf} 
\usepackage{xcolor}
\DeclareGraphicsExtensions{.pdf,.eps,.png,.jpg,.mps} 
\usepackage{hyperref}
\hypersetup{
    colorlinks=true,
    linkcolor=blue,
    citecolor=blue,
    filecolor=blue,      
    urlcolor=blue,
}

\RequirePackage{fancyhdr}  
\begin{document}
\title{Directly pumped 10 GHz microcomb modules from low-power diode lasers}


\author{Myoung-Gyun Suh$^{1}$, Christine Y. Wang$^{2}$, Cort Johnson$^{2}$, and Kerry Vahala$^{1*}$\\
$^1$T. J. Watson Laboratory of Applied Physics, California Institute of Technology, Pasadena, California 91125, USA.\\
$^2$Charles Stark Draper Laboratories, Cambridge, MA, 02139, USA\\
$^*$Corresponding author: vahala@caltech.edu
}

\maketitle\thispagestyle{fancy}
\newcommand{\ts}{\textsuperscript}
\newcommand{\tsb}{\textsubscript}

{\bf Soliton microcombs offer the prospect of advanced optical metrology and timing systems in compact form factors. In these applications, pumping of microcombs directly from a semiconductor laser without amplification or triggering components is desirable for reduced power operation and to simplify system design. At the same time, low repetition rate microcombs are required in many comb applications for interface to detectors and electronics, but their increased mode volume makes them challenging to pump at low power.  Here, 10 GHz repetition rate soliton microcombs are directly pumped by low-power ($<$ 20 mW) diode lasers. High-Q silica microresonators are used for this low power operation and are packaged into fiber-connectorized modules that feature temperature control for improved long-term frequency stability.}\\

Optical frequency combs (OFCs) have revolutionized a wide range of applications \cite{diddams2010evolving} and in recent years a miniature frequency comb or microcomb has been demonstrated using compact (often chip-based) microresonators \cite{kippenberg2018dissipative}. These devices use the optical Kerr effect to induce soliton mode locking from a continuous-wave optical pump \cite{herr2014temporal,brasch2016photonic,yi2015soliton,joshi2016thermally,wang2016intracavity}. Soliton microcombs provide a pathway towards miniaturization of OFC systems for application to time-standards \cite{newman2018photonic}, optical frequency synthesis \cite{spencer2018integrated}, precision spectroscopy \cite{suh2016microresonator,dutt2018chip}, telecommunications \cite{marin2017microresonator,mazur2018enabling}, LIDAR \cite{suh2018soliton,trocha2018ultrafast}, imaging \cite{bao2018microresonator}, and astronomy \cite{suh2018searching,obrzud2017astrocomb}. Pumping power of microcombs is an important design consideration and severely impacts generation of low repetition rate solitons. Such devices are required for microcomb interface to detectors and electronics, but are particularly challenging to operate at low pumping power. This can be understood by considering the Kerr parametric oscillation threshold power which is given by,
\cite{kippenberg2004kerr,yi2015soliton}
$$ P_{\text{th}} = \frac{n \omega_0 A_{\text{eff}}}{8 n_2} \frac{1}{\eta Q^2_{\text{T}}} \frac{1}{f_{\text{FSR}}} $$
where $n$ is refractive index, $\omega_0$ is the optical frequency, $A_{\text{eff}}$ is effective mode area, $n_2$ is the Kerr coefficient, $\eta = Q_{\text T}/Q_{\text E}$ is the waveguide-to-resonator loading factor, $Q_{\text T}$ ($Q_{\text E}$) is the total (external) quality factor, and $f_{\text{FSR}}$ is the free spectral range (FSR) or the repetition rate. The operating power of all microcombs is a multiple of this threshold power, and the expression makes clear that decreased repetition rate (smaller FSR) increases threshold and in turn comb power. Physically, this happens because the pumping volume of the resonator is larger for smaller FSR devices.

\begin{figure}[t!]
\centering
\includegraphics[width=\linewidth]{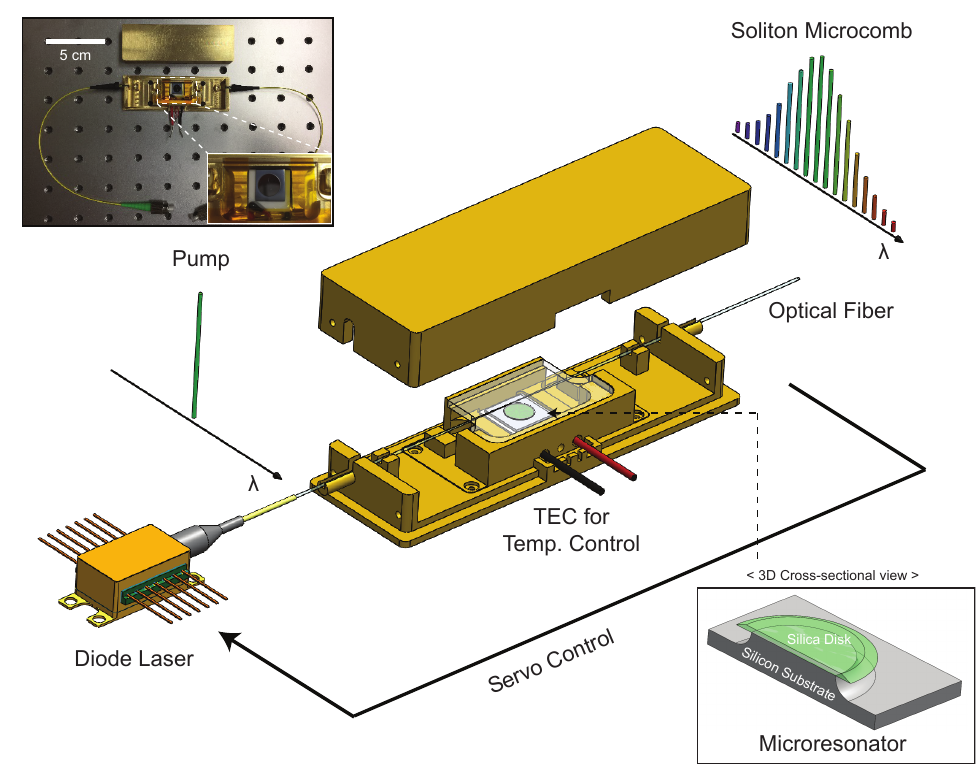}
\captionsetup{singlelinecheck=no, justification = RaggedRight}
\caption{{\bf Schematics of experimental setup and microcomb module.} Single wavelength diode laser directly pumps a silica microresonator packaged into a fiber-connectorized module to generate a 10 GHz soliton microcomb. Pump laser is polarization controlled before optical coupling via fiber taper. Output soliton power is tapped and used as an error signal for servo control of laser \cite{yi2016active}. Device temperature is controlled using a thermoelectric cooler (TEC). Upper left inset: photograph of 10 GHz soliton module with fiber pigtails. Lower right inset: schematic cross-sectional view of silica wedge disk resonator. Optical modes are guided at the perimeter of the wedged silica disk.}
\end{figure}

The impact of lower FSRs on microcomb power can be offset by leveraging the favorable inverse quadratic dependence of power on optical Q factor in the threshold equation as has been demonstrated using crystalline and silica-based microcombs \cite{herr2014temporal,yi2015soliton,suh2018gigahertz,yang2018bridging}. Recently, such designs have been applied to realize directly-pumped microcombs with low-repetition-rate (silica: 22 GHz, $\sim$ 40 mW \cite{volet2018micro}, and crystalline: 12.5 GHz, $>$ 100 mW pump \cite{pavlov2018narrow}), wherein a resonator is directly connected to a pump laser without the need for intermediate amplification or soliton triggering hardware.
It is also worth noting that improving optical Q factors in silicon nitride resonators have enabled recent demonstrations of $\sim$ 100 GHz soliton rates with $\sim$ 10 mW optical pumping \cite{liu2018ultralow} as well as $<$ 1 W electrical pumping \cite{raja2018electrically}; and $\sim$ 200 GHz soliton rates with $\sim$ 100 mW electrical pumping from a battery \cite{stern2018battery}.

\begin{figure}[t!]
\centering
\includegraphics[width=\linewidth]{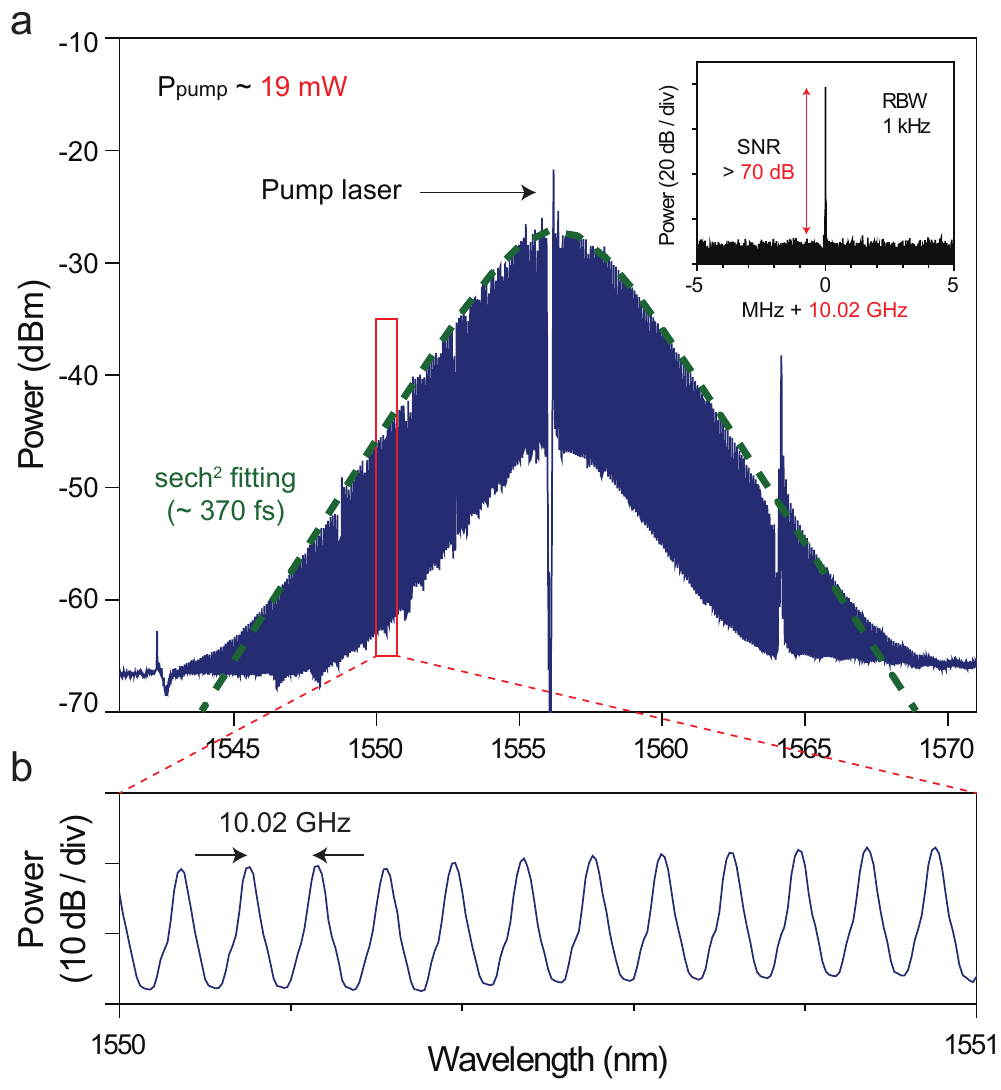}
\captionsetup{singlelinecheck=no, justification = RaggedRight}
\caption{{\bf Soliton spectrum.} (a) Optical spectrum of 10 GHz soliton microcomb at 19 mW pump power. The squared hyperbolic secant envelope (dashed green curve) gives a soliton pulse width of 370 fs. Pump laser line at 1556 nm is filtered out by a fiber bragg grating. Inset: Electrical spectrum showing the soliton repetition rate of 10.02 GHz. Signal-to-noise (SNR) is greater than 70 dB for a resolution bandwidth (RBW) of 1 kHz. (b) Zoom-in of the optical spectrum over a 1 nm span.}
\end{figure}

 In this letter, direct pumping of 10 GHz repetition rate soliton microcombs from low-power ($<$ 20 mW) diode lasers is reported. In the experiment, the high-Q whispering-gallery-mode silica resonators \cite{lee2012chemically} featured an $\sim$ 6.5 mm diameter corresponding to a 10.02 GHz soliton comb line spacing. Single soliton generation was possible at $\sim$ 15 mW pump power using the devices. Typical unloaded quality factors and parametric oscillation threshold for this design are reported elsewhere \cite{suh2018gigahertz}. The resonators are specifically designed for operation in the final frequency division stage of an optical atomic clock \cite{newman2018photonic} and a robust fiber-taper coupling inside a compact module (30 mm x 94 mm x 15 mm) was implemented (Fig.1) that included fiber pigtails and a thermoelectric cooling (TEC) element. Two modules were fabricated at the California Institute of Technology (Pasadena, CA, USA) and shipped to the Draper Laboratories (Cambridge, MA, USA) for testing relating to the clock experiment.

\begin{figure}[t!]
\centering
\includegraphics[width=\linewidth]{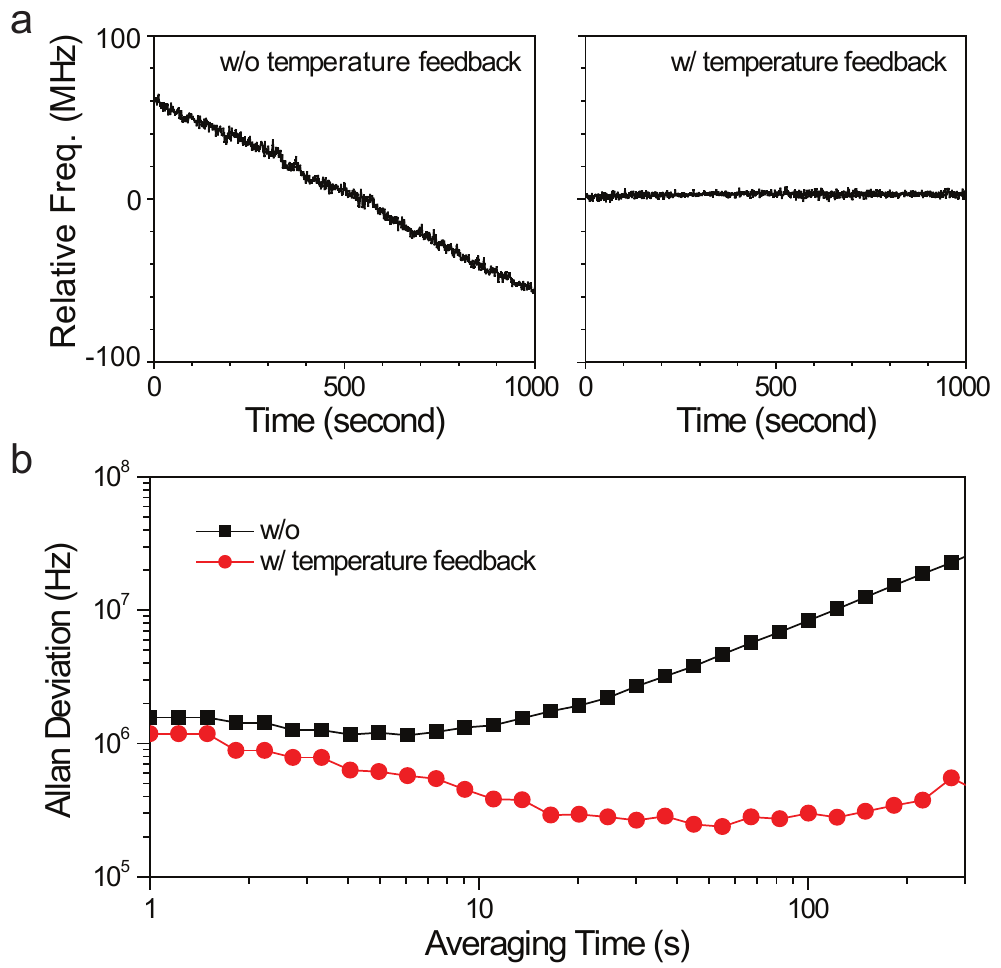}
\captionsetup{singlelinecheck=no, justification = RaggedRight}
\caption{{\bf Temperature feedback control of soliton offset frequency.} (a) Relative frequency drift of a comb line relative to an HCN reference laser without (left) and with (right) temperature feedback control. The data is acquired with 1 second gate time over 1000 seconds using a frequency counter. (b) Allan deviation plot of the data in panel (a) showing 300 kHz deviation at 100 seconds averaging.}
\end{figure}

In the experimental setup at Draper Laboratories, the packaged devices are directly pumped by a diode laser (RIO PLANEX Series 1556 nm laser). The diode laser has a short-term linewidth of 1 kHz and a maximum output power of approximately 23 mW at 150 mA bias current. Laser frequency can be swept at a speed of 100 GHz/s by tuning the bias current. This enables rapid tuning of the pump into the soliton existence detuning range for soliton triggering \cite{volet2018micro}. After the transmitted pump power is filtered using a fiber bragg grating (FBG) the soliton power is tapped via a 90/10 fiber splitter. The 10 percent port of soliton power is photodetected and used as an error signal for servo control to maintain the soliton state \cite{yi2016active}. Generation of single soliton states using under 20 mW pump power was possible in both devices. Fig. 2a shows the optical spectrum of the generated soliton from one device at 19 mW pump power. The squared hyperbolic secant envelope (dashed green curve) indicates the soliton pulse width is 370 fs. The electrical spectrum (Fig. 2a inset) shows the soliton repetition rate of 10.02 GHz with $>$ 70dB signal-to-noise. The zoomed-in optical spectrum in Fig. 2b shows the equidistantly-spaced comb lines.

To improve the long term stability of the system, the device temperature is controlled via a TEC element installed inside the module. In an independent measurement (performed at Caltech) to test the performance of the temperature feedback loop, one of the comb lines was heterodyned with an HCN reference laser (1559 nm Clarity laser manufactured by Wavelength References, Inc.) and the beat frequency was measured using a frequency counter. Figure 3a shows that the temperature feedback can suppress the beat frequency drift due to the environmental temperature changes. With the temperature feedback, an absolute frequency instability of 300 kHz at 100 second averaging time (Fig. 3b) was measured. Considering that the HCN reference laser shows an absolute frequency instability of $\sim$ 200 kHz at 100 seconds averaging, we can conclude that an absolute stability of the comb offset frequency better than 200 kHz is achieved at 100 seconds averaging with the temperature feedback.

In summary, 10 GHz soliton microcomb devices were fabricated, packaged into fiber connectorized modules, transported across the North American continent and then successfully operated by direct pumping from low-power ($<$ 20 mW) diode lasers. This low direct pumping power represents a record for this low comb repetition rate, and its X-band rate also enables use of low-bandwidth detectors and simplifies the electronic interface with the microcomb system (here a clock). These devices can be potentially integrated with SiN waveguides \cite{yang2018bridging} for more compact systems. The modules also feature a built-in TEC control that was able to stabilize the comb offset frequency in an open lab environment better than 200 kHz at 100 s averaging time.
\newline

\noindent
\textbf{Funding Information}

The authors acknowledge research funding from the Defense Advanced Research Projects Agency (DARPA) Atomic Clocks with
Enhanced Stability (ACES) program (HR0011-16-C-0118). The views, opinions
and/or findings expressed are those of the authors and should not be interpreted as representing the official views or policies of
the Department of Defense or the U.S. Government. This work is also supported by the Kavli Nanoscience Institute

\bibliography{main}


\end{document}